\documentclass[conference]{IEEEtran}
\IEEEoverridecommandlockouts
\usepackage{cite}
\usepackage{amsmath,amssymb,amsfonts}
\usepackage{algorithmic}
\usepackage{graphicx}
\usepackage{textcomp}
\usepackage{xcolor}
\def\BibTeX{{\rm B\kern-.05em{\sc i\kern-.025em b}\kern-.08em
    T\kern-.1667em\lower.7ex\hbox{E}\kern-.125emX}}
\begin{document}

% 10 + 2 => camera-ready 11 + 2
\title{Fork Entropy: Assessing the Diversity of Open Source Software Projects' Forks
%{\footnotesize \textsuperscript{*}Note: Sub-titles are not captured in Xplore and should not be used}
% \thanks{Identify applicable funding agency here. If none, delete this.}
}

% \author{\IEEEauthorblockN{Liang Wang, Zhiwen Zheng, Xiangchen Wu, Baihui Sang, and Xianping Tao}
% \IEEEauthorblockA{\textit{State Key Laboratory for Novel Software Technology}, 
% \textit{Nanjing University}, Nanjing, China \\
% \{wl,txp\}@nju.edu.cn, \{zwZheng,DZ21330024,jieruizhang\}@smail.nju.edu.cn, xchenwuhhu@gmail.com}
% }

\author{
\IEEEauthorblockN{Liang Wang}
\IEEEauthorblockA{\textit{State Key Laboratory for}\\ \textit{Novel Software Technology}, \\
\textit{Nanjing University}, Nanjing, China \\
wl@nju.edu.cn}
\\
\IEEEauthorblockN{Baihui Sang}
\IEEEauthorblockA{\textit{State Key Laboratory for}\\ \textit{Novel Software Technology}, \\
\textit{Nanjing University}, Nanjing, China \\
DZ21330024@smail.nju.edu.cn}

\and
\IEEEauthorblockN{Zhiwen Zheng}
\IEEEauthorblockA{\textit{State Key Laboratory for}\\ \textit{Novel Software Technology}, \\
\textit{Nanjing University}, Nanjing, China \\
zwZheng@smail.nju.edu.cn}
\\
\IEEEauthorblockN{Jierui Zhang}
\IEEEauthorblockA{\textit{State Key Laboratory for}\\ \textit{Novel Software Technology}, \\
\textit{Nanjing University}, Nanjing, China \\
jieruizhang@smail.nju.edu.cn}

\and
\IEEEauthorblockN{Xiangchen Wu}
\IEEEauthorblockA{\textit{State Key Laboratory for}\\ \textit{Novel Software Technology}, \\
\textit{Nanjing University}, Nanjing, China \\
xchenwuhhu@gmail.com}
\\
\IEEEauthorblockN{Xianping Tao}
\IEEEauthorblockA{\textit{State Key Laboratory for}\\ \textit{Novel Software Technology}, \\
\textit{Nanjing University}, Nanjing, China \\
txp@nju.edu.cn}
}

\maketitle

\begin{abstract}
On open source software (OSS) platforms such as GitHub, forking and accepting pull-requests is an important approach for OSS projects to receive contributions, especially from external contributors who cannot directly commit into the source repositories. Having a large number of forks is often considered as an indicator of a project being popular. While extensive studies have been conducted to understand the reasons of forking, communications between forks, features and impacts of forks, there are few quantitative measures that can provide a simple yet informative way to gain insights about an OSS project's forks besides their count. Inspired by studies on biodiversity and OSS team diversity, in this paper, we propose an approach to measure the diversity of an OSS project's forks (i.e., its fork population). We devise a novel fork entropy metric based on Rao's quadratic entropy to measure such diversity according to the forks' modifications to project files. With properties including symmetry, continuity, and monotonicity, the proposed fork entropy metric is effective in quantifying the diversity of a project's fork population. To further examine the usefulness of the proposed metric, we conduct empirical studies with data retrieved from fifty projects on GitHub. We observe significant correlations between a project's fork entropy and different outcome variables including the project's external productivity measured by the number of external contributors' commits, acceptance rate of external contributors' pull-requests, and the number of reported bugs. We also observe significant interactions between fork entropy and other factors such as the number of forks. The results suggest that fork entropy effectively enriches our understanding of OSS projects' forks beyond the simple number of forks, and can potentially support further research and applications.
\end{abstract}


\section{Introduction}\label{sec:introduction}

Forks play a central role in modern pull-based Open Source Software (OSS) development as precursors of making contributions to source repositories through pull-requests \cite{jiang2017and,chacon2014pro,li2020redundancy,gousios2014exploratory}.
On social coding platforms such as GitHub and GitLab, forks are created for various purposes including but not limited to archiving, learning, fixing bugs, adding new features \cite{mergel2015open,jiang2017and}.
Existing studies suggest that forks create increased opportunities for community engagement and voluntary participation \cite{zhou2019fork,dabbish2012social}.
Having a large number of forks is often considered as an indicator of an OSS project being popular \cite{dabbish2012social,vasilescu2015quality}.
However, merely having a large number of forks does not necessarily imply a productive and healthy OSS project.
Inefficient fork practices can lead to lost, rejected, or redundant contributions as pointed out in existing studies \cite{li2020redundancy,zhou2019fork}.
Understanding the factors related to an OSS project's fork efficiency can provide valuable insights to help project maintainers and contributors to diagnose and improve their fork practices \cite{zhou2019fork}. 

To gain insight into and improve forking efficiency, prior research has explored the relationship between a project's fork efficiency and its own characteristics, such as modularity and coordination \cite{zhou2019fork}. 
In contrast, this work takes a different perspective by examining the \textit{population of forks} created from a project. 
Specifically, we propose to measure the \textit{diversity of an OSS project's forks} by evaluating their differences in terms of modifications made to project files with the Rao's quadratic entropy---a measure widely adopted in quantifying population and ecological diversity \cite{rao1982diversity,botta2005rao}.
Our interest in understanding the diversity, beyond the simple count, of forks is inspired by the importance of diversity in different related domains:
1) \textit{Biodiversity} is shown to have significant effects on species' productivity, resilience, competition, and survival \cite{hughes2008ecological}.
In this work, we follow the promising approach of applying concepts and measures from ecological studies to the field of OSS research \cite{raja2012defining,jansen2014measuring}, and study how a project's fork population diversity is linked to its development.
2) \textit{OSS Teams Diversity}, including social diversity \cite{aue2016social}, gender and tenure diversity \cite{vasilescu2015gender,catolino2019gender}, culture and country diversity \cite{daniel2013effects,vasilescu2015perceptions}, and linguistic diversity \cite{vasilescu2013babel}, has been extensively studied and shown to be significantly linked to team performance.
While the above work focuses on the diversity of OSS contributors, our study examines the diversity of the artifacts they create, i.e., the forks.
3) Previous work in \textit{redundant change detection} \cite{ren2019identifying} and \textit{fork feature extraction} \cite{zhou2018identifying} has demonstrated how valuable information, such as patch content, changed file lists, clusters of features, etc, can be derived from forks' changes to project files. 
In line with this prior work, we focus on the diversity of forks regarding their modifications to project files.
To summarize, this study broadens the scope of diversity discussed in OSS development to the artifacts, i.e., forks, by introducing a new quantity called \textbf{fork entropy} to understand a project's fork population beyond simply the number of forks.
To the best of our knowledge, this work is among the first to quantify the diversity of an OSS project's fork population, and study its correlation with pull-based OSS development.

While anyone can fork from a public OSS repository, in this study, we specifically focus on forks and contributions from external contributors who cannot directly commit to OSS projects mainly for the following three reasons.
First, with the OSS model, anyone can potentially be an external contributor, resulting in a large population in this group \cite{padhye2014study}.
Second, external contributions' contribution through pull-requests and bug reports are critical during the development and maintenance of OSS projects \cite{gousios2014exploratory,krishnamurthy2016peripheral,vasilescu2015quality}.
Investigating external contributors' forks and contributions is crucial for understanding their effects on OSS projects' productivity and fork efficiency.
Third, fork and pull-request is the primary, if not the only, approach for external contributors to contribute since they do not have write access to the projects \cite{pinto2018challenges,padhye2014study}, making our study more focused on fork-related features and performance.
It is worth noting that while we restrict our investigation to external contributors' forks, we may also include forks owned by core members if the project strictly follows the pull-based development model, i.e., fork and contribute back to the source repository through pull-requests instead of directly pushing commits into the source repository, for all contributors including the core developers with write access.
We solely examine external contributors' forks in this study to maintain a focused study, and aim at answering the following research questions using data retrieved from fifty OSS projects' on GitHub.

\textbf{RQ1:} \emph{What is the correlation between an OSS project's fork entropy and its external productivity?}

In this work, we operationalize an OSS project's external productivity \cite{vasilescu2013stackoverflow,vasilescu2015gender,vasilescu2015quality,wang2020unveiling} as the number of commits integrated from external contributor-owned forks into the project's source repository.
The results of regression analysis suggest a project's fork entropy is significantly and positively correlated to its external productivity, especially for projects younger in ages.

\textbf{RQ2:} \emph{How does an OSS project's fork entropy relate to its acceptance rate of external pull-requests?}

The acceptance rate of external pull-requests serves as a useful proxy for measuring fork efficiency \cite{zhou2019fork} and can provide insight into a project's openness to external contributions \footnote{https://chaoss.community/kb/metric-change-request-acceptance-ratio/}. In this study, we operationalize the acceptance rate as the proportion of closed pull-requests raised by external contributors from their respective forks that were ultimately merged into the project. Our analysis indicates that fork entropy has a significant, positive correlation with the acceptance rate of external pull-requests, particularly when those pull-requests involve modifications to frequently changed `hot' files.

\textbf{RQ3:} \emph{What is the correlation between fork entropy and OSS projects' number of reported bugs?}

The number of reported bugs is commonly used as a proxy for evaluating software quality in projects \cite{khomh2012faster,ray2014large,vasilescu2015quality,foucault2015impact,wang2020unveiling}, and in this study, it is operationalized as the monthly reported bugs.
Our analysis reveals a significant, negative correlation between fork entropy and the number of reported bugs in OSS projects.
We also find that a high level of fork entropy is linked to a reduced rising trend of bug-reporting issues with the increasing number of forks in OSS projects.

Through the above studies, we conclude that the diversity of OSS projects' fork population plays an important role in understanding the development of OSS projects under the pull-based model.
And the proposed fork entropy metric is an effective and useful indicator in assessing fork diversity, which shows significant correlations to key productivity and quality indicators about OSS development and maintenance.
In summary, this work makes the following contributions.
\begin{itemize}
    \item We propose to measure the diversity of OSS projects' fork populations to gain insights about pull-based OSS development, which provides a novel point of view about the forks and diversity about OSS projects.
    \item We devise the fork entropy metric based on Rao's quadratic entropy, and show properties of the metric including symmetry, continuity, and monotonicity, which makes the metric effective in measuring fork diversity.
    \item We conduct empirical studies to reveal the correlations between fork entropy and the external productivity, the acceptance rate of external pull-requests, and number of reported bugs in OSS projects, which demonstrate the usefulness in understanding the diversity of forks.
    \item We discuss the implications of fork entropy in understanding and guiding practices of pull-based social coding for OSS development.
\end{itemize}

The rest of this paper is organized as follows.
Sec. \ref{sec:related_work} introduces the background and related work.
Sec. \ref{sec:measure_of_fork_diversity} presents the process of calculating the proposed fork entropy metric and shows its properties.
In Sec. \ref{sec:methods}, we present the design of empirical studies, and report the results in Sec. \ref{sec:results}.
We discuss the implications of this work, and the threats to validity in Sec. \ref{sec:discussion}.
Finally, we conclude the paper in Sec. \ref{sec:conclusion}.

\section{Background \& Related Work}
\label{sec:related_work}

This section introduces the background and related work.

\subsection{Studies on Forks and Pull-Based OSS Development}

The pull-based development model is a modern paradigm for software development that is particularly well-suited to geographically distributed teams \cite{gousios2014exploratory}. 
This workflow can be broken down into seven steps, which include: forking, cloning, editing, syncing, pushing, submitting, and evaluating \cite{chacon2014pro,li2020redundancy}. 
By providing a code base and a set of tools for task management, code review, and DevOps, the pull-based model simplifies participation and lowers entry barriers for external contributors compared to traditional patch-based models \cite{gousios2014exploratory}. 

Forks can be created for many different reasons or intentions.
In \cite{jiang2017and}, Jiang et al. summarizes developers' reasons and preferences of creating forks, and OSS projects' attracting characteristics of getting forked.
They discover various reasons of forking including contributing back with pull-requests, fixing bugs, adding new features, keeping copies, etc.
St{\u{a}}nciulescu et al. conduct a case study using the Marlin project to explore the reasons, benefits, are challenges of forks \cite{stuanciulescu2015forked}.
According to whether they will contribute back to the source repository, forks can be classified as ``hard'' forks \cite{hadian2022exploring,zhou2020has}, independently developed forks (IDFs) \cite{rastogi2016forking}, or divergent forks \cite{businge2022reuse} that are not intended to be merged back, and ``social'' forks \cite{zhou2020has} that are created to make contributions.

\begin{figure*}[t!]
	\centering
	\includegraphics[width=\linewidth]{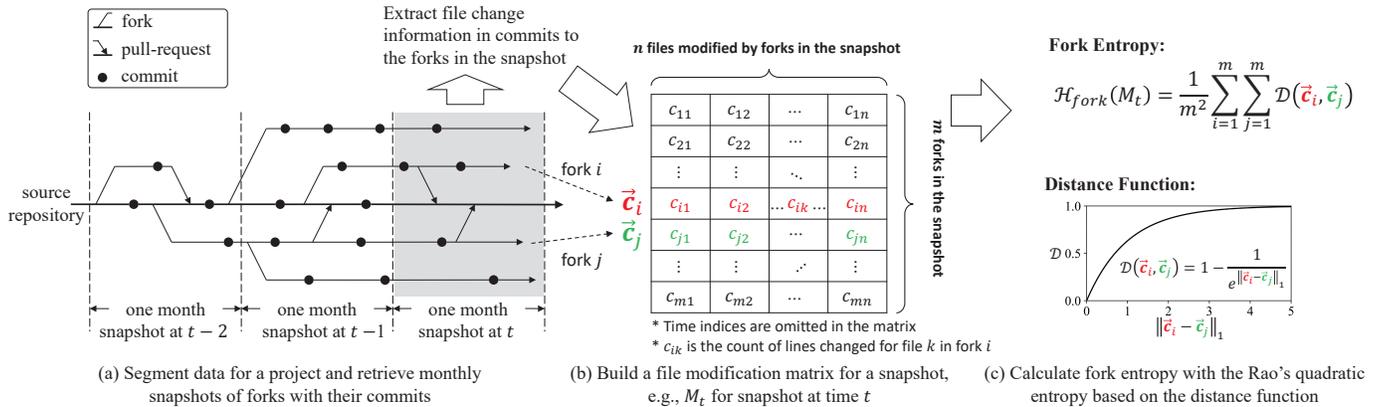}
	\caption{Overall process of calculating fork entropy.}
	\label{fig:measure_of_fork_diversity}
\end{figure*}

The many forks created from a source repository (with different intentions above) form the fork population studied in this work.
The concept is closely related to the concept of software family in existing work \cite{brisson2020we,hadian2022exploring,businge2022reuse}.
In \cite{brisson2020we}, Brisson et al. study the communication in a software family with respect to following, pull-requests, and issues, and analyze the correlation between such communication and a project's star counts.
In \cite{businge2022reuse}, Businge et al. explore the characteristics including the size, package dependencies, categories, etc, of software families, as well as their code propagation practices.
Software families composed of hard forks are studied in the above work \cite{brisson2020we,businge2022reuse}.
In \cite{hadian2022exploring}, Hadian et al. study the evolution and communication between repositories in a family of projects (forks).
They discover differences between hard and social forks in terms of activity, deviation of dependencies, and communication.
In this work, we study the diversity of a projects' contributors-owned fork population created on GitHub.
Although we do not discriminate hard and social forks in our studies, we are more interested in contributing forks \cite{rastogi2016forking} by analyzing the correlation between the forks' diversity and contributions received in the source repository.

For owners of contributing forks that follow the pull-based model, they can fork and develop in their own forked repositories before contributing back to the source repository through pull-requests.
However, coordination issues such as misalignment and conflict among developers can lead to inefficiencies in forking practices, which include lost contributions, rejected pull requests, redundant development, and fragmented communities as summarized by Zhou et al in \cite{zhou2019fork}. 
Consequently, it is essential to understand the factors, and develop tools to aid efficient forking practices.
In \cite{zhou2019fork}, Zhou et al. explore characters of the source repository such as modularity, and coordination mechanisms on the efficiency of forking practices.
To better understand the forks of OSS projects, tools are developed to visualize source code changes in forks \cite{imamura2022bug}, and identify features from commits in a projects forks \cite{zhou2018identifying}.
In this work, we propose a new metric called fork entropy to facilitate studies of the diversity of a forks based on their modification to files, which offers a new perspective to understand an OSS project's fork population.

\subsection{Studies on External Contributors and Pull-Requests}

In this work, we focus on forks owned, and contributions made by external contributors because we focus on the diversity and contributions related to OSS projects' forks.
Unlike core developers who have write access, and can directly commit into a source repository, external contributors make contributions to projects through patches or pull-requests \cite{padhye2014study,gousios2014exploratory}.
In \cite{padhye2014study}, Padhye et al. categorize code committers into core, external, and mutant, and find that the number of external committers of projects developed by popular scripting languages is comparable with the number of core committers.
After inspecting the pull-requests opened in 2013 in GitHub,  Gousios et al. conclude that pull-requests is a way of projects to get external contributions because 73.07\% of the pull-requests have been merged using facilities provided by GitHub \cite{gousios2014exploratory}.
In \cite{vasilescu2015quality}, Vasilescu et al. study the influence of adopting continuous integration (CI) on the of acceptance of core and external (non-core) developers' pull-requests, as well as the impact on software quality.
They discover an improvement in projects' ability in integrating external contributions without sacrificing code quality after adopting CI.
By inverstigating the Eclipse community, Sinha et al. discover factors, such as demonstration of knowledge and skill in bug repositories, can influence the promotion of external developers to core committers\cite{sinha2011entering}.
In brief, contributions made by external contributors through pull-requests play an important role in OSS development and maintenance which worth studying.
Moreover, external contributors are a suitable group for our study on fork diversity and its correlation on OSS contributions  because, unlike core developers with write access, they need to fork a repository before making contributions.

\subsection{Studies on Diversity Related to OSS Projects}
Most existing software engineering studies related to our work focus primarily on the diversity of team members in OSS projects. Gender diversity has been widely examined and found to have positive effects on project growth \cite{aue2016social}, team productivity \cite{vasilescu2015gender}, and community health \cite{catolino2019gender}. Studies have also reported a positive correlation between the country diversity of team members and project growth \cite{aue2016social}, and the impact of tenure diversity on team productivity has been suggested \cite{vasilescu2015gender}. Additionally, Daniel et al. measured the reputation and role diversity of participants and found positive effects on community engagement and market success \cite{daniel2013effects}. However, as far as we know, there is currently a dearth of formal measures of fork diversity. Moreover, the diversity metrics presented above are typically measured using either the Blau index \cite{blau1977inequality} or the coefficient of variation \cite{allison1978measures}. While the Blau index is well-suited for measuring diversity in categorical variables (such as gender \cite{vasilescu2015gender} and country \cite{aue2016social}), and the coefficient of variation can measure dispersion in numerical variables (like tenure and reputation), neither method is ideal for calculating the proposed metric of fork entropy due to limitations on the forms of data associated with understanding the fork populations of OSS projects.

Diversity is widely acknowledged as a multifaceted concept, which can be categorized into three main components: richness, evenness, and disparity \cite{jost2006entropy}. Richness refers to the absolute number of species in a population, evenness describes the distribution of species abundances, and disparity denotes the differences among species \cite{daly2018ecological}. Due to its complexity, diversity has been operationalized using different metrics, depending on the specific application. For instance, while species richness is a commonly used indicator of diversity that emphasizes the richness component, the Gini coefficient, which measures income inequality, highlights the disparity component \cite{daly2018ecological}. In the context of OSS development, forks may naturally differ due to differences in expertise, experience, and intentions among developers \cite{scacchi2006understanding}. Therefore, we focus on the disparity component of diversity and quantify the differences in file modifications among forks. To this end, we employ Rao's quadratic entropy, which measures the expectation of dissimilarity between two samples randomly taken from a population \cite{rao1982diversity}. This approach has been widely used in studies of population and ecological diversity \cite{rao1982diversity,botta2005rao}. Further details on how we calculate fork entropy can be found in Section \ref{sec:measure_of_fork_diversity}.

\section{Fork Entropy}
\label{sec:measure_of_fork_diversity}
This section presents the proposed measure of fork diversity with fork entropy.

\subsection{Overview}\label{sec:steps}

Fig. \ref{fig:measure_of_fork_diversity} illustrates the overall process of calculating fork entropy to measure the diversity of forks created for an OSS project.
As shown in Fig. \ref{fig:measure_of_fork_diversity}(a), for each project involved in our study, we first collect data about its forks, and construct a series of snapshots of \emph{fork populations} with predefined time intervals.
We select projects from GitHub for easy data collection.
We empirically set the time interval to one month in this work after consulting the literature \cite{wang2020unveiling}.
Second, as shown in Fig. \ref{fig:measure_of_fork_diversity}(b), we build a \emph{file modification matrix} to contain the changes made by the forks in snapshot at time $t$, denoted as $M_t$.
To keep the notations concise, we slightly abuse the notations to omit time index $t$ when referring to the file modification matrix for a snapshot in the rest of the paper.
Finally, as shown in Fig. \ref{fig:measure_of_fork_diversity}(c), we calculate the fork entropy with the Rao's quadratic entropy defined on the pair-wise \emph{distance function} that quantifies the difference between each pair of forks given a file modification matrix.
Detailed steps are as follows.

\subsection{Construct Fork Populations in Snapshots}

The first step is to construct the fork population as shown in Fig. \ref{fig:measure_of_fork_diversity}(a).
Given an OSS project, we first locate its source repository, e.g., \verb|tensorflow/tensorflow|, in the database.
We then perform a breadth first search using the `forked\_from' key in the database to retrieve the direct forks and indirect forks (i.e., `forks of forks') of the source repository in an iterative manner until all forks of the project are retrieved.
The retrieved forks are segmented by time intervals with a fixed duration (empirically set to a month in this work) to get a series of snapshots.
Each snapshot contain forks with their commits received during the time interval.
For a fork to be included in a snapshot, it must: 1) have at least one commit with file modifications during the snapshot's time interval; and 2) is owned by a external contributor who does not have write access to, or have privileges to close issues or pull-requests in the source repository \cite{zhou2019fork}.
We determine a contributor to be external if he / she has never directly commit into a repository, or performed any privileged actions such as closing issues opened by other users.
A fork can be included in multiple snapshots if it meets the above criteria in different time intervals, but with different commits.
We restrict our scope to forks owned by external members of the project to conduct a focused study as introduced in the beginning of the paper.
It should be noted that fork ownership is not a part of the proposed fork entropy metric.
Following the above approach, the set of forks in each snapshot forms a \textit{fork population} of the project during the corresponding time interval.

% More specifically, we use breadth first search and retain a queue that initially contains only the source repository.
% In each round, we first pop up all repositories in the queue.
% For each repository that popped up, we query the direct forks of the repository from GHTorrent and push them into the queue.
% We repeat the above process until no new fork is pushed into the queue and gather all forks that have stayed in the queue as the candidate set.
% For each fork in the candidate set, we subsequently remove it if any following condition is met: the fork 1) does not change in the given month; or 2) is created and owned by a core developer of the source repository.
% We make clear that a fork changes if it pushes at least one commit consisting of any file modification, and a core developer is who either had write access to the source repository or had closed issues or pull-requests of others in the source repository \cite{zhou2019fork}.
% After the filter of the candidate set, we get the fork population consisting of forks created by external developers and have changed in the given month.

\subsection{Build Fork File Modification Matrix for Each Snapshot}

Next, we build a \textit{file modification matrix} to contain the changes made by the fork population in a snapshot.
In the matrix, a fork, e.g., the $i$-th fork, is encoded as a row vector: $$\vec{\mathbf{c}}_i = \langle c_{i1},c_{i2}, \cdots, c_{ij}, \cdots, c_{in} \rangle^\top,$$
where $n$ is the number of files in the project that are modified by one of the forks in the population, and $c_{ij}\in \mathbb{R}$ is the count of lines in file $j$ which are modified by fork $i$ during the time interval of the snapshot.
Intuitively, $\vec{\mathbf{c}}_i$ is the fingerprint of the $i$-th fork in terms of modifications to project files.
Let $m$ be the number of forks in the population, we can obtain the \textit{file modification matrix}, $M \in \mathbb{R}^{m \times n}$, by stacking the row vectors of all of the $m$ forks as shown in Fig. \ref{fig:measure_of_fork_diversity}(b).
Because the set of files modified by fork populations in different snapshots are likely to be different, it is common that the number of columns, $n$, varies with different time intervals.
We include only files that are modified by the fork population in a snapshot to guarantee that the file modification matrix does not contain rows or columns that are all zeros.

\subsection{Calculate Fork Entropy with Rao's Quadratic Entropy}

For each snapshot, we use the Rao's quadratic entropy with a distance function defined on the file modification matrix $M$ to calculate the average degree of difference between forks in the population following 
Eq. (\ref{eq:quadratic_entropy}).
\begin{equation}
\begin{footnotesize}
    \mathrm{QE}(M)=\frac{1}{m^{2}}\sum_{i=1}^{m}\sum_{j=1}^{m}\mathcal{D}(\vec{\mathbf{c}}_{i},\vec{\mathbf{c}}_{j}),
\label{eq:quadratic_entropy}
\end{footnotesize}
\end{equation}
where $\mathrm{QE}(M)$ denotes the Rao's quadratic entropy that estimates the expectation of difference between two individuals randomly selected from the population \cite{rao1982diversity}, $M$ is the file modification matrix correspond to the snapshot (with the time index $t$ omitted), $m$ is the number of forks in the population, $\vec{\mathbf{c}}_{i},\vec{\mathbf{c}}_{j}$ are the the $i$- and $j$-th row in $M$, respectively, and $\mathcal{D}$ is a distance function that quantifies the degree of difference between two vectors as defined in Eq. (\ref{eq:distance}) and visualized in Fig. \ref{fig:measure_of_fork_diversity}(c).
\begin{equation}
\begin{footnotesize}
    \mathcal{D}(\vec{\mathbf{c}}_{i},\vec{\mathbf{c}}_{j}) = 1-\exp(-\gamma\|\vec{\mathbf{c}}_{i}-\vec{\mathbf{c}}_{j}\|_{1}),
\label{eq:distance}
\end{footnotesize}
\end{equation}
where $\exp(-\gamma\|\vec{\mathbf{c}}_{i}-\vec{\mathbf{c}}_{j}\|_{1})$ is the Laplacian kernel \cite{rupp2015machine}, $\|\cdot\|_{1}$ is the 1-norm, and $\gamma$ is the hyperparameter used to adjust the sensitivity of the function to differences.
We adopt the Laplacian kernel because it is a non-linear transform sensitive to slight change and performs excellently in many detection tasks, e.g., character recognition \cite{fadel2016investigating}.
% {\color{red} what is the intuitive name for gamma in nature language? the power? suggest to mark gamma in the equation, and say in the text that gamma = 1}.
% The reason for us to adopt the distance function is to {\color{red} shrink the value range xxx you mean control or limit the range of the distance? in what interval?} and balance the distribution of differences {\color{red} what does balance the distribution mean?}.
% After the negative exponential transform visualized in Fig. \ref{fig:measure_of_fork_diversity}(c), the larger differences (e.g., large than 5) are strictly scaled to near 1, while the smaller differences (e.g., less than 5) are loosely scaled in the range from 0 to 1 {\color{red} generally know what you are talking about, but hard to understand}.

Practical considerations also led us to adopt the Laplacian kernel in our study. Our analysis revealed that many forks in our dataset contain only minor changes to project files---some forks modifying only a single line in a single file, similar to findings from previous research \cite{alali2008s,arafat2009commit}. Consequently, we observed a substantial number of small differences, resulting in a fork entropy distribution that roughly followed a bell curve. After testing various distance functions, including the Laplacian and Gaussian kernels \cite{rupp2015machine}, we ultimately selected the distance function presented in Equation (\ref{eq:distance}) based on its real-world performance.

By substituting Eq. (\ref{eq:distance}) into Eq. (\ref{eq:quadratic_entropy}), we have the Eq. (\ref{eq:fork_diversity}) for fork entropy.
\begin{equation}
\begin{footnotesize}
    \mathcal{H}_{fork}(M)=\frac{1}{m^{2}}\sum_{i=1}^{m}\sum_{j=1}^{m}\Big(1-\exp(-\gamma\|\vec{\mathbf{c}}_{i}-\vec{\mathbf{c}}_{j}\|_{1})\Big),
\label{eq:fork_diversity}
\end{footnotesize}
\end{equation}
where $\gamma$ is set to 1 to compute the raw difference between two vectors in practice.
With the above definition, it is trivial to see that $0 \leq \mathcal{H}_{fork} < 1$, and $\mathcal{H}_{fork}$ takes the minimum value when the numbers of changed lines for all files modified by different forks are identical.

\subsection{Properties of the Proposed Fork Entropy}\label{sec:construct_validity}

%We next evaluate the construct validity of fork entropy on measuring the diversity of forks in the context of pull-based OSS development.
Considering the basic axioms \cite{daly2018ecological} of a diversity index and expectations in the particular context jointly, the following properties are met by the proposed fork entropy:

\textbf{Symmetry.} \textit{The fork entropy is not related to the order of forks during its calculation.}
It is straightforward to see $\mathcal{H}_{fork}$ satisfies the symmetry property because its distance function $\mathcal{D}$ is symmetric, i.e., $\mathcal{D}(\vec{\mathbf{c}}_{i},\vec{\mathbf{c}}_{j}) = \mathcal{D}(\vec{\mathbf{c}}_{j},\vec{\mathbf{c}}_{i})$.

\textbf{Continuity.} \textit{The fork entropy is a continuous function.} 
It is also easy to see that $\mathcal{H}_{fork}$ is in a continuous interval $\mathcal{H}_{fork}\in[0,1)$ by its definition.

Symmetry and continuity are two fundamental properties of a diversity index \cite{daly2018ecological}.
Next, we introduce monotonicity as a property to meet our goal of measuring fork diversity.

\textbf{Monotonicity.} \textit{Adding a redundant (or distinctive) fork will decrease (or increase) fork entropy.}
We first explain the terms before proving monotonicity.
Given $m$ existing forks and a new fork $\vec{\mathbf{c}}_{m+1}$, Eq. (\ref{eq:conflict}) calculates the difference of the new fork to existing ones.
\begin{equation}
\begin{footnotesize}
    \widetilde{\mathcal{D}}(\vec{\mathbf{c}}_{m+1})=\frac{1}{m}\sum_{i=1}^{m}\mathcal{D}(\vec{\mathbf{c}}_{i},\vec{\mathbf{c}}_{m+1}).
\label{eq:conflict}
\end{footnotesize}
\end{equation}
We say $\vec{\mathbf{c}}_{m+1}$ is \emph{redundant} if its difference to existing forks is less than the average difference among the existing $m$ forks; in contrast, $\vec{\mathbf{c}}_{m+1}$ is \emph{distinctive} if the new fork's difference to existing forks exceeds the average difference among the existing $m$ forks.

Assuming that the vector for a new fork $\vec{\mathbf{c}}_{m+1}$ is added to an existing file modification matrix $M$ that contains $m$ forks to obtain a new matrix $M'$, we derive the new fork entropy of $M'$ in Eq. (\ref{eq:monotonicity_1}).
\begin{equation}
\begin{footnotesize}
\begin{aligned}
    \mathcal{H}_{fork}(M')&=\frac{1}{(m+1)^{2}}\sum_{i=1}^{m+1}\sum_{j=1}^{m+1}\mathcal{D}(\vec{\mathbf{c}}_{i},\vec{\mathbf{c}}_{j})\\
    &=\frac{1}{(m+1)^{2}}\Bigg(\sum_{i=1}^{m}\sum_{j=1}^{m}\mathcal{D}(\vec{\mathbf{c}}_{i},\vec{\mathbf{c}}_{j})+2\sum_{i=1}^{m}\mathcal{D}(\vec{\mathbf{c}}_{i},\vec{\mathbf{c}}_{m+1})\Bigg)\\
    &=\frac{m^{2}}{(m+1)^{2}}\mathcal{H}_{fork}(M)+\frac{2}{(m+1)^{2}}\sum_{i=1}^{m}\mathcal{D}(\vec{\mathbf{c}}_{i},\vec{\mathbf{c}}_{m+1}).
\end{aligned}
\label{eq:monotonicity_1}
\end{footnotesize}
\end{equation}

We denote $\mathcal{H}_{fork}(M')-\mathcal{H}_{fork}(M)$ as $\Delta$ and obtain Eq. (\ref{eq:monotonicity_2}) by substituting $\Delta$ into Eq. (\ref{eq:monotonicity_1}).
\begin{equation}
\begin{footnotesize}
\begin{aligned}
    \Delta&=\frac{2m+1}{(m+1)^{2}}\Bigg(\frac{1}{m+0.5}\sum_{i=1}^{m}\mathcal{D}(\vec{\mathbf{c}}_{i},\vec{\mathbf{c}}_{m+1})-\mathcal{H}_{fork}(M)\Bigg)\\
    &\approx\frac{2m+1}{(m+1)^{2}}\Bigg(\widetilde{\mathcal{D}}(\vec{\mathbf{c}}_{m+1})-\mathcal{H}_{fork}(M)\Bigg).
\end{aligned}
\label{eq:monotonicity_2}
\end{footnotesize}
\end{equation}

According to Eq. (\ref{eq:monotonicity_2}), $\Delta$ is negative when $\vec{\mathbf{c}}_{m+1}$ is redundant and is positive when $\vec{\mathbf{c}}_{m+1}$ is distinctive.
It means that fork entropy decreases after adding a redundant fork while increases after adding a distinctive fork.
Consequently, fork entropy possesses the above properties and is valid to quantify the diversity of forks.

\section{Methods for Empirical Studies}
\label{sec:methods}

This section presents the variables, dataset, and analysis methods for the studies about the proposed fork entropy with respect to our research questions.

\subsection{Variables}

We measure a project's outcomes, including external productivity, the acceptance rate of external pull-requests, and number of reported bugs.
We also introduce control variables relevant to those outcomes.

\textbf{Outcome: external productivity.}
The number of commits is a widely used indicator of the productivity of OSS projects \cite{vasilescu2013stackoverflow,vasilescu2015gender,vasilescu2015quality,wang2020unveiling}.
In this work, we focus on external productivity that quantifies external contributor' contributions to a project, which is measured by the number of commits integrated into the project's source repository through pull-requests from external contributor-owned forks.
The monthly external productivity is obtained to perform regression analysis with fork entropy in each snapshot to answer \textbf{RQ1}.
A possible threat lies in that maintainers may change the origins of commits, for example, by ``cherry-picking'' in pull-requests \cite{gousios2014exploratory}, which can cause us to miss some contributions made by external contributors. 
Fortunately, we find such cases are rare in our dataset after manual inspections.

It is important note that software developers' productivity is a multifaceted concept which covers the activity, performance, efficiency, satisfaction and well-being, etc, of the developers as suggested in \cite{forsgren2021space}, and can be influenced by many factors beyond the ones studied in this paper such as the team sizes \cite{muric2019collaboration}.
In this paper, we adopt the number of commits as a commonly used measure of productivity with respect to developers' activity, which should not be considered as a comprehensive measure of external developers' productivity.

\textbf{Outcome: external pull-request acceptance rate.}
Researchers regard the acceptance rate of pull-requests as a crucial indicator of the development efficiency of OSS projects \cite{zhou2019fork} because maintainers reject pull-requests that are obsolete, conflicting, duplicated, etc \cite{gousios2014exploratory,steinmacher2018almost,nadri2020insights}.
We focus on the acceptance rate of external pull-requests delivered from external contributor-owned forks to an OSS project's source repository, measured as the proportion of the merged pull-requests among closed ones.
The list of merged and closed pull-requests can be obtained from GitHub's rest API.
The monthly acceptance rate of external pull-requests is calculated to perform regression analysis with fork entropy in each snapshot to answer \textbf{RQ2}.
As many developers integrate pull-requests via other mechanisms rather than GitHub interface, the status of pull-requests is not very reliable reported by GitHub \cite{gousios2014exploratory,zhou2019fork}.
We follow the heuristics first proposed by Gousios et al. \cite{gousios2014exploratory} and subsequently refined by Zhou et al. \cite{zhou2019fork} to determine pull-requests' status.
A pull-request is considered been merged if any of the following conditions is met: 
1) Maintainers perform a `merged' action for the pull-request on GitHub. 
2) The pull-request is closed by a commit using certain phrase conventions (e.g., \texttt{fixes $\#$1234}) advocated by GitHub\footnote{https://github.blog/2011-04-09-issues-2-0-the-next-generation/}, or any of the last three comments of the pull-request refers to a commit indicating the merge of the pull-request\footnote{Comment matches regular expression \texttt{(merg$|$apply$|$appl$|$pull$|$ push$|$integrat$|$land$|$cherry(-$|$$\backslash$s+)pick$|$squash)(ing$|$i?ed)}.}, and the commit exists in the source repository's commit history.

\textbf{Outcome: number of reported bugs.}
The number of bugs per unit time is a popular proxy of code quality \cite{khomh2012faster,ray2014large,vasilescu2015quality,foucault2015impact,wang2020unveiling}.
We refer to \cite{vasilescu2015quality,wang2020unveiling} to count emerging bug-report issues in each snapshot.
We do not distinguish bug reports from core members or external contributors.
To identify bug-report issues, we process issue titles and labels by lowercasing and Porter stemming \cite{willett2006porter} then search bug-related keywords, including \textit{defect}, \textit{error}, \textit{bug}, \textit{issue}, \textit{mistake}, \textit{incorrect}, \textit{fault}, and \textit{flaw}.
If the title or any label of an issue contains at least one keyword, we mark it as a bug-report issue.
The monthly code quality is assessed to perform regression analysis with fork entropy in each snapshot to answer \textbf{RQ3}.
Since the count-based assessment of code quality relies heavily on the issue base, a threat arises if projects rarely utilize GitHub's default issue-trackers.
Thus, we examine the number of issues in each project and exclude projects with few issues.

\textbf{Control variables.}
Based on prior software engineering literature \cite{gousios2014exploratory,tsay2014influence,vasilescu2015quality,zhou2019fork} and our experience, we include the following factors potentially relevant to the above project outcomes as control variables.
\begin{itemize}
	\item \textbf{NumForks} and \textbf{NumFiles}: The two variables are the number of forks and that of modified files, respectively. They jointly describe the shape of a file modification matrix. The more forks a project has, the more pull-requests are submitted by non-core developers \cite{vasilescu2015quality}.
	\item \textbf{ProjectAge}: The age of a project's source repository in days. The older the project, the fewer external pull-requests maintainers merge or reject \cite{vasilescu2015quality}.
	\item \textbf{NumStars}: The number of stars a project's source repository receives. This variable usually refers to the popularity of OSS projects. External contributors are more likely to contribute to more popular projects \cite{vasilescu2015quality}.
	\item \textbf{RatioOldContributors}: The ratio of external contributors with prior experience in successfully submitting pull-requests to a project's source repository. Core developers prefer to trust contributors they have worked with before \cite{gousios2014exploratory,tsay2014influence,zhou2019fork}.
	\item \textbf{RatioPRsWithTests}: The ratio of pull-requests that contain test cases. A pull-request has test cases if any file pathname contains `test' \cite{tsay2014influence}. Pull-requests that contain test cases are more likely to be merged \cite{tsay2014influence}.
	\item \textbf{RatioPRsTouchHotFiles}: The ratio of pull-requests that touch hot files. A hot file is that one modified by any merged pull-request in the past three months \cite{gousios2014exploratory}. Existing studies suggest pull-requests that modify hot files are more likely to be accepted \cite{gousios2014exploratory}.
\end{itemize}

\subsection{Data Collection}

We collect data through GHTorrent \cite{Gousi13} and GitHub REST API\footnote{https://docs.github.com/en/rest}, with the aid of the OSS Compass community \cite{oss-compass}.
We start by selecting the most popular five thousand projects from the May 2019 GHTorrent dump according to the number of stars a source repository receives.
Then, we filter projects based on the following criteria.
\begin{itemize}
	\item \textit{Projects that do not develop software applications or frameworks are removed.} We remove projects that serve for document storage or course teaching. We examine project names and `README' files by searching keywords, including \textit{awesome}, \textit{homework}, \textit{assignment}, \textit{course}, \textit{note}, and \textit{document}. If any keyword is found, we remove the project after manually rechecking. We also delete projects with no programming-language-specific files by looking at the file extensions.
	\item \textit{Projects whose number of active forks or external pull-requests are less than one hundred are removed.} To ensure a project has sufficient forks and contributions, we retain projects that 1) contain at least one hundred active forks, i.e., forks that have pushed at least one commit into the forked repository \cite{businge2022reuse}, and 2) contain at least one hundred external pull-requests submitted by external contributors.
	\item \textit{Projects whose issues are less than one hundred are also removed.} To reduce the threat to \textbf{RQ3}, we conservatively exclude projects with less than one hundred issues to ensure the remaining projects actively use the default issue-trackers on GitHub.
\end{itemize}

Among the 2533 candidate projects selected with the above criteria, we sample fifty projects\footnote{A complete list of the fifty projects studied in this paper can be found at https://github.com/wangliang-cs/fork-entropy-ase-2023-repos.} to cover different application domains, which include ten application software (e.g., \verb|Atom/atom|), two system software (e.g., \verb|kubernetes/kubernetes|), sixteen web libraries and frameworks (e.g., \verb|angular/angular.js|), nine non-web libraries and frameworks (e.g., \verb|tensorflow/tensorflow|), and thirteen software tools (e.g., \verb|Microsoft/vscode|).

\subsection{Regression Analysis}\label{sec:regression_analysis}

We calculate monthly variables for each project from its creation to May 2019.
All variables in each project snapshot compose an independent unit for regression analyses.
We omit units with empty fork populations because fork entropy is meaningless without forks, even if it has a value of zero.
Before fitting models with the data, we perform log-transform on control variables under skewed distributions to stabilize variance and reduce heteroscedasticity \cite{metz1978basic}, including \emph{NumForks}, \emph{NumFiles}, \emph{NumStars}, and \emph{RatioOldContributors}.
% Some predictor variables are on very different scales: consider rescaling
Then, we standardize fork entropy and all control variables to make the mean of each one is zero and the standard deviation is one, which makes all estimated coefficients of the model are on the same scale.
Additionally, we manually examine distributions of outcome variables and conservatively remove about 1\% of values as outliers to ensure the models are robust against outliers \cite{osborne2004power}.

We build generalized linear mixed models (GLMMs) to analyze the correlations between fork entropy and the outcome variables.
GLMMs inherit from generalized linear models (GLMs) to allow response variables from non-normal distributions and extend GLMs to include both fixed and random effects \cite{breslow1993approximate}.
After exploratory data analysis, GLMMs are appropriate because the response variables in our data are non-normal and show apparent variability among projects.
Specifically, fork entropy, control variables, and interactions between fork entropy and each control variable are modeled as fixed effects.
To capture the project-to-project variability in the response (e.g., some projects naturally attract more external contributors and receive more contributions than others), we add a random-effects term for projects into the models.
We also allow for deviations in the slope of the number of a project's forks from the population values (i.e., we accept the possibility that, for example, projects with higher initial external productivity may, on average, be less strongly affected by the increase in fork counts).
We use the \verb|glmer| function provided by the \verb|lme4| package \cite{2016Linear} in \verb|R| to build models.
Following prior practices \cite{gelman2006data,zhou2019fork}, the Poisson and logistic regressions are specified for \emph{count} (i.e., external productivity and number of reported bugs) and \emph{ratio} (i.e., the acceptance rate of external pull-requests) response variables, respectively.
We also explicitly set the denominator (i.e., the number of closed pull-requests) from the ratio as the \verb|weights| parameter when modeling the acceptance rate of external pull-requests.

We check the collinearity among independent variables using the variance inflation factors (VIF below five is recommended \cite{cohen2014applied}).
All values are below 2 in our models, which means collinearity is not a problem in our data.
We adopt the marginal R-squared ($R_{m}^{2}$) and the conditional R-squared ($R_{c}^{2}$) to assess the goodness-of-fit of the models.
$R_{m}^{2}$ and $R_{c}^{2}$ describe the proportion of variance explained by the fixed effects alone and explained by the fixed and random effects together, respectively \cite{johnson2014extension,nakagawa2013general}.
In addition, we report the estimated effect, standard error, and significance level (i.e., $p$-value) for each model variable.
We also report the results of ANOVA type-\uppercase\expandafter{\romannumeral2} analysis, and Akaike's information criteria (AIC) \cite{akaike1998information} and Bayesian's information criteria (BIC) \cite{schwarz1978estimating} for each model using the \verb|car| and \verb|performance| package \cite{daniel2021performance} in \verb|R|, respectively.
%A variable's effect size is the proportion of deviance explained by the model that can be attributed to the variable.

\section{Results of the Studies}
\label{sec:results}

This section shows the results of regression analyses for our research questions.

\subsection{RQ1: What is the correlation between an OSS project's fork entropy and its external productivity?}

\begin{table}[t!]
	\centering
	\caption{Fixed effects of external productivity model.}
	\label{table:external_productivity_model}
	\begin{tabular}{llc}
		\hline
		& \multicolumn{1}{c}{Estimate (Std. Errors)} & \multicolumn{1}{c}{Chisq} \\ \hline
		(Intercept)                      & \textcolor{white}{-}3.360 (0.136)***                  &                             \\
		$\mathcal{H}_{fork}$                    & \textcolor{white}{-}0.325 (0.005)***                & 5038.09***                    \\
		NumForks                         & \textcolor{white}{-}0.614 (0.075)***                & \textcolor{white}{00}67.84***                    \\
		NumFiles                         & \textcolor{white}{-}0.510 (0.006)***                & 7700.90***                    \\
		ProjectAge                       & -0.176 (0.005)***               & 1097.94***                    \\
		NumStars                         & \textcolor{white}{-}0.032 (0.004)***                & \textcolor{white}{00}98.00***                     \\
		RatioOldContributors               & \textcolor{white}{-}0.079 (0.003)***                & \textcolor{white}{0}600.72***                     \\
		$\mathcal{H}_{fork}$:NumForks           & -0.116 (0.006)***               & \textcolor{white}{0}383.30***                     \\
		$\mathcal{H}_{fork}$:NumFiles           & \textcolor{white}{-}0.064 (0.005)***                   & \textcolor{white}{0}147.55***                        \\
		$\mathcal{H}_{fork}$:ProjectAge         & -0.123 (0.004)***               & 1070.85***                    \\
		$\mathcal{H}_{fork}$:NumStars           & \textcolor{white}{-}0.087 (0.005)***                  & \textcolor{white}{0}311.58***                       \\
		$\mathcal{H}_{fork}$:RatioOldContributors & \textcolor{white}{-}0.015 (0.003)***                  & \textcolor{white}{00}21.59***                        \\ \hline
		\multicolumn{3}{l}{AIC=65465.38; BIC=65558.12; $R_{m}^{2}$=0.40; $R_{c}^{2}$=0.98}   \\
		\multicolumn{3}{l}{Num. obs.=3579; Num. groups: ProjectID=50}                                        \\ \hline
		\multicolumn{3}{l}{*** $p<0.001$, ** $p<0.01$, * $p<0.05$}  \\
		\multicolumn{3}{l}{Log-transformed and standardized variables following Sec. \ref {sec:regression_analysis}.}  
	\end{tabular}
\end{table}

% \begin{figure}[t!]
% 	\centering
% 	\includegraphics[width=\linewidth]{figs/coefplot_1.eps}
% 	\caption{Estimated coefficients and 95\% confidence intervals from the external productivity model in Table \ref{table:external_productivity_model}.}
% 	\label{fig:coefplot1}
% \end{figure}

To answer \textbf{RQ1}, we model the number of commits integrated from external contributor-owned forks into a project's source repository as a function of fork entropy.
In the regression, we control factors including the number of forks, the number of modified files, project age, the number of stars, and the ratio of external contributors with prior experience.
Table \ref{table:external_productivity_model} summarizes the regression results.
We can see that the model is effective by explaining about 98\% of the variance with the fixed and random effects together, which exceeds the fixed effects alone by approximately 58 percentage points.

The results in Table \ref{table:external_productivity_model} suggest that fork entropy significantly and positively correlates to the external productivity of OSS projects ($p$-value below 0.001), with the third largest estimated coefficient just lower than the coefficients of NumForks and NumFiles.
This result suggest that projects with more diverse fork populations integrate more commits created by external contributors.
%Furthermore, by looking at the effect size shown by the `Chisq' column in Table \ref{table:external_productivity_model}, we find that fork entropy is an important relating factor to external contributors' productivity because it contributes to 30.5\% of the deviance explained by the model.
In addition, the controlled factors in the model are also significantly linked to external productivity.
Table \ref{table:external_productivity_model} shows that all controls except project age are positively correlated to the external productivity of OSS projects.
As a result, OSS projects with more forks and modified files, higher popularity, younger in age, and more proportion of external contributors with prior experience generally correspond to higher external productivity.

\begin{figure}[t!]
	\centering
	\includegraphics[width=\linewidth]{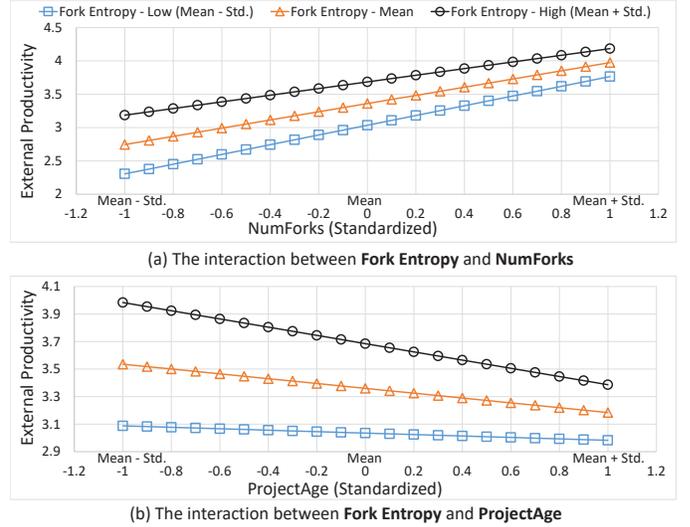}
	\caption{Interactions in the external productivity model shown in Table \ref{table:external_productivity_model}.}
	\label{fig:interactions4rq1}
\end{figure}

We next analyze the interactions between fork entropy and the control variables in the model.
As shown in Table \ref{table:external_productivity_model}, all interactions show significance.
The interactions between fork entropy and the number of forks and between fork entropy and project age negatively correlate to the external productivity of OSS projects, and the remaining interactions show positive correlation with the response.
We can also observe from the table that the two interactions negatively correlated to the response have the two largest effects as discussed below.

Fig. \ref{fig:interactions4rq1}(a) illustrates the trends of external productivity with the increasing number of forks at low, middle, and high levels of fork entropy, respectively.
We define the low, middle, and high levels of fork entropy respectively to correspond to its values with mean minus one standard deviation (Mean - Std.), mean, and mean plus one standard deviation (Mean + Std.) after consulting the literature \cite{aiken1991multiple}.
In general, external productivity raises with the increase in NumForks.
And the interaction suggests a higher level of fork entropy is associated with a lower growth rate of external productivity with respect to NumForks.
With a fixed number of forks, an increase in a project's fork entropy is related to an increased external productivity.
This result suggests that, a project with a larger and more diverse population of forks have a higher probability of receiving  contributions from the external contributors.%, which results in a more productive and viable project.

Fig. \ref{fig:interactions4rq1}(b) depicts the interaction between fork entropy and project age.
We can see that the external productivity generally decreases with the increasing of project age.
The ratio of decreasing in external productivity with increased project age is amplified with higher levels of fork entropy.
A possible explanation to this result is, younger projects under active development are possibly more open to accepting diverse contributions from external contributors than older projects.
And projects older in age may have possibly entered a stage of stable maintenance, or slowly dying for lacking the capacity of handling diverse external contributions.

Fork entropy shows a positive and significant interaction between each of the other control variables including NumFiles, NumStars, and RatioOldContributors, respectively, as reported in Table \ref{table:external_productivity_model}.
We omit the details due to page limits.

In summary, we answer \textbf{RQ1} as follows.
\textit{Fork entropy shows a significant, positive correlation with the external productivity of OSS projects. Further analysis uncovers significant interactions between fork entropy and other variables, including the number of forks, project age, number of files, number of stars, and ratio of old contributors. We observe that increasing fork entropy is related to a higher level of external productivity, particularly with respect to the number of forks. Additionally, our results indicate that for younger projects, increasing fork entropy shows a stronger positive relation with external productivity compared to older projects.}

\subsection{RQ2: How does an OSS project's fork entropy relate to its acceptance rate of external pull-requests?}

\begin{table}[t!]
	\centering
	\addtolength{\tabcolsep}{-4pt}
	\caption{Fixed effects of external pull-request acceptance rate model.}
	\label{table:acceptance_efficiency_model}
	\begin{tabular}{llc}
		\hline
		& \multicolumn{1}{c}{Estimate (Std. Errors)} & \multicolumn{1}{c}{Chisq} \\ \hline
		(Intercept)                        & -0.729 (0.195)***                   &                             \\
		$\mathcal{H}_{fork}$                      & \textcolor{white}{-}0.179 (0.015)***                    & \textcolor{white}{0}145.07***                       \\
		NumForks                           & -0.196 (0.158)                     & \textcolor{white}{000}1.55\textcolor{white}{***}                        \\
		NumFiles                           & \textcolor{white}{-}0.049 (0.020)*                     & \textcolor{white}{00}11.54***                       \\
		ProjectAge                         & \textcolor{white}{-}0.490 (0.011)***                    & 2710.74***                     \\
		RatioOldContributors                 & \textcolor{white}{-}0.614 (0.010)***                    & 4578.80***                     \\
		RatioPRsWithTests                  & \textcolor{white}{-}0.135 (0.017)***                    & \textcolor{white}{00}63.97***                       \\
		RatioPRsTouchHotFiles               & -0.091 (0.015)***                    & \textcolor{white}{00}43.30***                       \\
		$\mathcal{H}_{fork}$:NumFiles           & -0.122 (0.012)***                   & \textcolor{white}{00}97.07***                       \\
		$\mathcal{H}_{fork}$:ProjectAge           & -0.053 (0.009)***                   & \textcolor{white}{00}38.38***                       \\
		$\mathcal{H}_{fork}$:RatioOldContributors   & -0.028 (0.009)**                   & \textcolor{white}{00}10.01**\textcolor{white}{*}                       \\
		$\mathcal{H}_{fork}$:RatioPRsWithTests    & -0.230 (0.011)***                   & \textcolor{white}{0}421.33***                      \\
		$\mathcal{H}_{fork}$:RatioPRsTouchHotFiles & \textcolor{white}{-}0.140 (0.011)***                    & \textcolor{white}{0}155.37***                      \\ \hline
		\multicolumn{3}{l}{AIC=46173.12; BIC=46272.04; $R_{m}^{2}$=0.09; $R_{c}^{2}$=0.53}     \\
		\multicolumn{3}{l}{Num. obs.=3579; Num. groups: ProjectID=50}                                          \\ \hline
		\multicolumn{3}{l}{*** $p<0.001$, ** $p<0.01$, * $p<0.05$}                      \\
		\multicolumn{3}{l}{Log-transformed and standardized variables following Sec. \ref {sec:regression_analysis}.}                        
	\end{tabular}
\end{table}

% \begin{figure}[t!]
% 	\centering
% 	\includegraphics[width=\linewidth]{figs/coefplot_2.eps}
% 	\caption{Estimated coefficients and 95\% confidence intervals from the external pull-request acceptance rate model in Table \ref{table:acceptance_efficiency_model}.}
% 	\label{fig:coefplot2}
% \end{figure}

To answer \textbf{RQ2}, we study the correlation between fork entropy and the acceptance rate of pull-requests submitted by external contributors.
Here we calculate fork entropy based on a variant of the file modification matrix that only includes modifications involved in pull-requests.
We perform the adjustment because pull-requests allow us to figure out which modifications are submitted to source repositories in each time interval.
We exclude modifications made in forks but not included in pull-requests because they are largely invisible to maintainers who make decisions.
We apply logistic regression for the acceptance rate by considering fork entropy.
We control other factors, including NumForks, NumFiles, ProjectAge, RatioOldContributors, RatioPRsWithTests, and RatioPRsTouchHotFiles.
Interactions between fork entropy and each control variable are also involved in the model.
We ignore the interaction between fork entropy and NumForks because the control variable does not show a significant correlation with the response.
Table \ref{table:acceptance_efficiency_model} summarizes the model's results.
The model explains about 53\% of variability by including the fixed and random effects together, achieving a significant improvement compared to the fixed effects alone.

From Table \ref{table:acceptance_efficiency_model}, we find fork entropy significantly and positively correlates to the acceptance rate of external pull-requests, despite that fork entropy has a smaller estimated effect of 0.179 compared with the estimated coefficients of  RatioOldContributors, ProjectAge, and NumForks of 0.614, 0.49, -0.196, respectively.
However, the effect of NumForks fails to show statistical significance.
Except for RatioPRsTouchHotFiles, the other four control variables are positively linked to the acceptance rate of external pull-requests.

For the model's interaction terms, only the interaction between fork entropy and the proportion of pull-requests that touch hot files show significant and positive correlation with the acceptance rate of external pull-requests.
The remaining interactions are significantly and negatively correlated to the response.
Furthermore, we analyze the interactions with relatively notable effects, i.e., the interaction between fork entropy and RatioPRsTouchHotFile, as well as its interaction with RatioPRsWithTests, as follows.

\begin{figure}[t!]
	\centering
	\includegraphics[width=\linewidth]{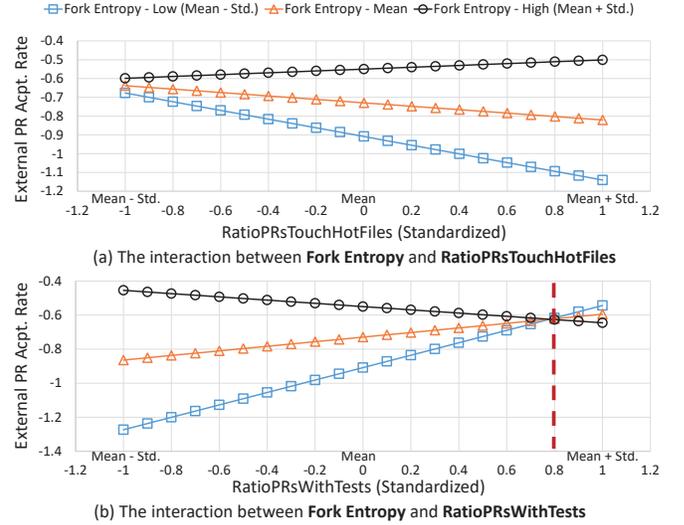}
	\caption{Interactions in the acceptance rate model shown in Table \ref{table:acceptance_efficiency_model}. Negative response values caused by the logit function.}
	\label{fig:interactions4rq2}
\end{figure}

First, as illustrated in Fig. \ref{fig:interactions4rq2}(a), the correlation between the proportion of pull-requests that touch hot files (RatioPRsTouchHotFiles) and external pull-requests' acceptance rate differs with different levels of fork entropy.
The acceptance rate declines with RatioPRsTouchHotFiles when fork entropy is at a median or low level, and increases when fork entropy is high.
A possible explanation is that, when fork entropy is at the median or low level, pull-requests are touching the same or similar sets of hot files, which result in conflicts and duplication among them, leading to a reduced acceptance rate.
On the contrary, when fork entropy is at a high level, the pull-requests are touching different, less overlapped, hot files in the recent history.
As a result, an increase in RatioPRsTouchHotFiles also leads to a increase in acceptance rate.
%The above observation is consistent with our intuition on the correlation between fork diversity and the efficiency of pull-based OSS development.

Next, Fig. \ref{fig:interactions4rq2}(b) illustrates the interaction between fork entropy and RatioPRsWithTests on the acceptance rate of external pull-requests.
We observe that an increase in the proportion of pull-requests that contain test cases is correlated to a higher acceptance rate when fork entropy is at a low level, which is consistent with existing observations that including test cases are helpful to make a pull request being accepted \cite{tsay2014influence}.
However, the above direction of correlation inverses at a high level of fork entropy.
From the perspective of varying fork entropy by fixing the other factor, i.e., comparing the left and right parts separated by the vertical dashed line in Fig. \ref{fig:interactions4rq2}(b), the acceptance rate generally increases with an increasing fork entropy with respect to the number of pull-requests contain test cases.
But when the majority of pull-requests contain test cases, as shown by the part to the right of the dashed line, the increase in fork entropy instead shows a negative correlation with the acceptance rate.

We omit the details about the interactions between fork entropy and other control variables including NumFiles, ProjectAge, and ProjectAge, which show significant and negative interactions as listed in Table \ref{table:acceptance_efficiency_model} due to page limites.

In summary, our answer to \textbf{RQ2} is as follows.
\textit{Fork entropy has a statistically significant, positive correlation with the acceptance rate of external pull-requests, albeit with a limited effect. Our analysis indicates that fork entropy also positive interacts with the number of pull-requests that touch hot files on the response. We find that the acceptance rate of pull-requests touching hot files only exhibits an upward trend when fork entropy is at a high level.}

\subsection{RQ3: What is the correlation between fork entropy and OSS projects' number of reported bugs?}

\begin{table}[t!]
	\centering
	\caption{Fixed Effects of number of reported bugs model.}
	\label{table:code_quality_model}
	\begin{tabular}{llc}
		\hline
		& \multicolumn{1}{c}{Estimate (Std. Errors)}  & \multicolumn{1}{c}{Chisq}  \\ \hline
		(Intercept)                & \textcolor{white}{-}2.632 (0.204)***                        &                              \\
		$\mathcal{H}_{fork}$              & -0.086 (0.006)***                    & 222.49***                      \\
		NumForks                   & \textcolor{white}{-}0.356 (0.066)***                     & \textcolor{white}{0}32.51***                      \\
		NumFiles                   & -0.065 (0.007)***                    & \textcolor{white}{0}82.53***                      \\
		ProjectAge                 & \textcolor{white}{-}0.078 (0.007)***                     & 118.94***                      \\
		NumStars                   & \textcolor{white}{-}0.115 (0.006)***                     & 332.64***                      \\
		$\mathcal{H}_{fork}$:NumForks     & -0.063 (0.008)***                        & \textcolor{white}{0}62.36***                         \\
		$\mathcal{H}_{fork}$:NumFiles     & \textcolor{white}{-}0.045 (0.007)***                        & \textcolor{white}{0}40.92***                         \\
		$\mathcal{H}_{fork}$:ProjectAge   & \textcolor{white}{-}0.032 (0.005)***                      & \textcolor{white}{0}37.03***                        \\
		$\mathcal{H}_{fork}$:NumStars     & \textcolor{white}{-}0.037 (0.006)***                     & \textcolor{white}{0}38.55***                       \\ \hline
		\multicolumn{3}{l}{AIC=35052.62; BIC=35133.00; $R_{m}^{2}$=0.08; $R_{c}^{2}$=0.96} \\
		\multicolumn{3}{l}{Num. obs.=3579; Num. groups: ProjectID=50}                                    \\ \hline
		\multicolumn{3}{l}{*** $p<0.001$, ** $p<0.01$, * $p<0.05$}         \\
		\multicolumn{3}{l}{Log-transformed and standardized variables following Sec. \ref {sec:regression_analysis}.}                         
	\end{tabular}
\end{table}

% \begin{figure}[t!]
% 	\centering
% 	\includegraphics[width=\linewidth]{figs/coefplot_3.eps}
% 	\caption{Estimated coefficients and 95\% confidence intervals from the number of reported bugs model in Table \ref{table:code_quality_model}.}
% 	\label{fig:coefplot3}
% \end{figure}

In this section, we study the correlation between fork entropy and the number of reported bugs of OSS projects.
We model the number of bug-report issues as a function of fork entropy with controlling other factors including the number of forks, the number of modified files, project age, and the number of stars.
Interactions between fork entropy and each controlled factor are also involved in the model as fixed effects.
Table \ref{table:code_quality_model} summarizes the results of the number of reported bugs model.
The model fits the data well by explaining about 96\% of the variability by including both fixed and random effects, achieving a considerable improvement compared to the fixed effects alone by about 88 percentage points.

The results in Table \ref{table:code_quality_model} suggest that fork entropy significantly and negatively correlates to the number of bug-report issues with an estimated coefficient of -0.086.
The absolute value is the third largest among the factors, lower than the estimated coefficient of NumForks and ProjectAge, which are 0.356, and 0.115, respectively.
The results also suggest that the number of modified files significantly and negatively correlates to the number of bug-reporting issues, while the other controlled factors show significant and positive correlations with the response.
All interaction terms show significance, with the interaction of fork entropy and the number of forks negatively correlates to the response.

\begin{figure}[t!]
	\centering
	\includegraphics[width=\linewidth]{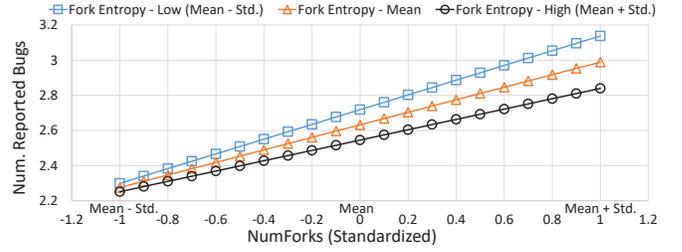}
	\caption{The interaction between fork entropy and NumForks in the number of reported bugs model shown in Table \ref{table:code_quality_model}.}
	\label{fig:interactions4rq3}
\end{figure}

We observe an opposition between fork entropy and the number of forks, where the former negatively correlates to the response and the latter positively relates to the response, respectively.
Fig. \ref{fig:interactions4rq3} shows the interaction.
We find that more bug-report issues are submitted with the increasing in the number of forks regardless of fork entropy.
There is a slower growth in bug-report issues when fork entropy is at a higher level.
Combined with the results in in \textbf{RQ 1} and \textbf{RQ 2}, a possible explanation to the above result is that increased fork entropy are related to more commits and pull-requests been accepted, including bug-fixing ones, before potential bugs are reported. 
With the above results, if we agree with the assumption that less reported bugs indicates a higher software quality \cite{khomh2012faster,wang2020unveiling}, we can state that a higher level of fork entropy is positively linked to an improved software quality.

In summary, we answer \textbf{RQ3} as below.
\textit{Our study reveals a significant, negative correlation between fork entropy and the number of reported bugs in OSS projects. Furthermore, it shows that the rising trend of bug-reporting issues associated with the increasing number of OSS projects' forks is reduced when fork entropy is at high levels.}

\section{Discussion}
\label{sec:discussion}

This section discusses the implications of our work, and the threats to validity.

\subsection{Implications and Discussion}

This section discuss the implications of our work.

Developing metrics and models that measure the health of OSS projects has received much attention from both researchers and practitioners of OSS recently.
For example, communities such as CHAOSS \cite{chaoss} and OSS Compass \cite{oss-compass} are founded to achieve an understanding, and provide services to measure the health of OSS communities, projects, and ecosystems through numerous qualitative and quantitative metrics.
Fork related metrics, such as the number of technical forks \cite{chaoss-metric-technical-fork}, is considered as a useful indicator.
The fork entropy proposed in this paper is a new metric about an OSS project's forks, and can potentially be added to the collection of metrics provided by the above communities due to its correlations with projects' external productivity, pull-request acceptance ratio, and number of reported bugs as shown in our studies.

% \textbf{For OSS projects' contributors and maintainers.}
% The inflow and retention of external contributors are critical for the sustainability and success of OSS projects \cite{zhou2016inflow}, making it a pressing issue for maintainers. However, the loose coordination inherent in the pull-based model can impede cooperation among project participants. For example, redundant efforts and mismatches with maintainers' intentions can lead to inefficiencies in pull-based social coding \cite{zhou2019fork}. To enhance coordination among developers, various mechanisms and tools have been proposed, such as claiming work in progress in issue-trackers\footnote{https://github.com/dear-github/dear-github/issues/191} on GitHub and increasing awareness of code changes in other forks \cite{zhou2018identifying}. Despite these efforts, there remains a need to establish a quantitative approach to comprehensively understand and diagnose the status of an OSS project's forks and assess the effectiveness of interventions aimed at improving forking efficiency.
% The proposed fork entropy fills the niche of quantifying the diversity of an OSS project's forks, which is potentially useful in fulfilling the above requirement.

%\textbf{Further research on OSS fork diversity.}

The proposed  fork entropy metric also provides opportunities for further research on OSS projects' forks. 
For instance, one could utilize pattern mining and time series analysis technologies to examine the evolving patterns and future trends of fork entropy over time \cite{amin2012automated}. 
Furthermore, it is potentially important to identify and assess the various factors and events that contribute to the rise and fall of fork diversity in OSS projects, and analyze their impact on the sustainability and prosperity of OSS projects \cite{raja2012defining,rastogi2016forking}. 
With a thorough understanding of fork diversity trends and influencing factors, it is possible to develop monitoring tools and guidelines to facilitate effective forking and collaboration during the development and maintenance of OSS projects.

\subsection{Threats to Validity}
We discuss the threats to validity as follows.

\textbf{Construct Validity.}
This work explores the diversity of OSS projects' fork populations measured by the proposed fork entropy metric.
The construct validity concerns about whether the fork entropy measures the diversity of the fork population.
Because fork entropy is built on top of the well-established metric of Rao's quadratic entropy \cite{rao1982diversity} which has shown to be effective in measuring population diversity \cite{botta2005rao}, and because we have shown the properties of for entropy including symmetry, continuity, and monotonicity, we argue that the proposed fork entropy is valid in measuring the diversity of an OSS project's fork population.

\textbf{Internal Validity.}
In this work, we conduct a data-driven approach to study the correlation between fork entropy and OSS projects' external productivity, pull-request acceptance rate, and number of reported bugs.
We also restrict our scope to forks owned, and contributions made, by external contributors.
The restriction is designed to make our study more focused on fork-related properties about OSS projects.
However, there are many other factors affecting the contributions made to an OSS project \cite{jansen2014measuring}.
External contributors' contribution are also partially determined by the core-members of OSS teams.
As a result, the conclusions made in this paper do not fully explain how fork-related contributions made to OSS projects or imply a causal relationship.
In addition, we set the time interval to one month when taking snapshots to build file modification matrices and calculate fork entropy.
Although this size is widely used in previous OSS-related studies \cite{wang2020unveiling}, changing the time granularity during the analysis may potentially change the results.
It is our future work to test the results with different time intervals.

\textbf{External Validity.}
We select fifty projects from GitHub that cover various application domains in our studies. 
However, the collection of projects studied is small compared to the number of projects hosted on OSS platforms. 
The conclusions made in this paper may not generalize well to other projects, or to projects hosted on other platforms. 
It is our future work to include more projects in our studies.

\section{Conclusion}
\label{sec:conclusion}

In this work, we focus on the pull-based OSS development and propose a novel metric called fork entropy to measure the diversity of the population of forks around an OSS project.
We calculate fork entropy by applying the Rao's quadratic entropy with a distance function that measures the dissimilarity of the forks' modifications to project files.
By conducting empirical studies on fifty real-world OSS projects from GitHub, we reveal that there exist significant correlations between fork entropy and the external productivity, the acceptance rate of external pull-requests, and number of reported bugs in these projects.
We also find significant interactions between fork entropy and other influencing factors such as the number of forks, project age, ratio of pull-requests that touch hot files, etc, which have previously found to correlate to the productivity and quality of OSS projects.
The proposed fork entropy metric not only enriches the current available metrics about understanding OSS projects' forks, it also offers opportunities for conducting further research on pull-based social coding.

\section*{Acknowledgment}

We thank OSS Compass and GHTorrent in helping us to obtain data about OSS projects' forks. This work is supported by NSFC No. 62172203, Fundamental Research Funds for the Central Universities, and the Collaborative Innovation Center of Novel Software Technology and Industrialization.

% The preferred spelling of the word ``acknowledgment'' in America is without 
% an ``e'' after the ``g''. Avoid the stilted expression ``one of us (R. B. 
% G.) thanks $\ldots$''. Instead, try ``R. B. G. thanks$\ldots$''. Put sponsor 
% acknowledgments in the unnumbered footnote on the first page.

\bibliographystyle{IEEETran}
\bibliography{IEEEabrv,references}

\end{document}